# HPC-Enabled Generator Importance Assessment for RTO-Scale Resource Adequacy Planning

Eve Tsybina, Nicholson Koukpaizan, Joshua Hambrick, Slaven Peles
Oak Ridge National Laboratory, Oak Ridge, TN, USA
Email: {tsybinae, koukpaizannk, hambrickjc, peless} @ ornl.gov
Samim Konjicija
University of Sarajevo, Faculty of Electrical Engineering,
Sarajevo, Bosnia and Herzegovina
Email: skonjicija@etf.unsa.ba

***Abstract* — Modern power systems are increasingly under stress as aging assets approach retirement and load growth outpaces new generation construction. The severity of this challenge varies by region: in the EU, the transmission grid can partially compensate for local generation shortfalls, while in the US, generation tends to be more localized, making retirements harder to offset. Retirement of generation has consequences for system reserves, fuel supply chain, and public health. We present an high-performance computing (HPC) framework for rapidly assessing the grid importance of individual generating units and ranking them by primary fuel type, operating cost, or grid impact. Historically, such studies were computationally intensive and therefore conducted infrequently. This work demonstrates that such assessments can be completed in minutes, enabling planners to evaluate a much broader range of generation portfolio scenarios than was previously possible.**

***Index Terms*—high-performance computing, generation ranking, generation retirement, fuel mix.**

## I. Introduction

Generation portfolios and generation adequacy are a central concern for the US and EU power sectors. In the EU, an increase in inverter-based generation following the European Green Deal has been the main driver of declining fossil generation, with inverter-based resources rising from a share of 34% in 2019 to 47% in 2024 [1], causing growing stress on the power system. According to the European Commission, a series of energy crises in 2022 and in 2026, prompted the region to reduce dependence on fossil fuels and caused a significant reassessment of the European generation portfolio [2]-[3]. Further, the problem is exacerbated by growing transmission needs [4]. More than half of the transmission projects needed by 2030 are still awaiting permits, and inverter-based projects totaling 1,700 GW across 16 European countries were in connection queues in 2024 [5]. Supply chain issues for fossil fuels have similarly caused concerns over resource adequacy, reinforcing the need for systematic approaches to generation portfolio management on both continents.

In the US, the growth of new load has significantly outpaced generation additions, causing stakeholders to defer retirement of generating units. The 2025 NERC Long-Term Reliability Assessment [6] highlights intensifying resource adequacy risks throughout the North American bulk power system, with summer peak demand forecast to grow by 224 GW over the next decade – a more than 69% increase over prior projections, driven largely by new data centers for artificial intelligence and the digital economy. PJM, one of the largest grid operators in the world, has warned that its system could face a capacity shortage as soon as the 2026/2027 delivery year, prompting emergency filings to expedite generation interconnection and address near-term supply-demand imbalances [7]. The 2026 NERC Long-Term Reliability Assessment [8] states that over 105 GW in peak seasonal capacity is announced for retirement over 2026-2036. Determining which units can be retired, which should be retained, and in what order, requires a systematic assessment of each generator's contribution to system reliability and cost. A recent meeting by FERC [9] further underscored these concerns, noting that PJM and MISO each had a net -10 GW of accredited capacity in 2015-2024.

Managing this challenge requires careful prioritization of the existing generation fleet. The decision to retire a given generating unit carries consequences not only for system reliability and operating cost, but also for the overall fuel mix – with downstream implications for emissions and public health that vary significantly depending on which plants are retired and in what order [10]. The historically high computational cost of physics-based modeling, such as transmission-constrained AC optimal power flow (ACOPF), has caused planners to rely on simplified or infrequent assessments. ACOPF is the gold standard for grid optimization, capturing the full non-linear AC power flow physics, but at system scale with thousands of generators and buses it is computationally intensive, and evaluating thousands of retirement scenarios sequentially has been prohibitive for routine planning cycles. Prior work on parallelizing optimal power flow problems has demonstrated significant speedups on multi-core hardware [11], and GPU-accelerated methods have shown further promise for large-scale applications [12], but the transition to truly exascale execution has only recently become practical.

[1]This manuscript has been authored by UT-Battelle, LLC under Contract No. DE-AC05-00OR22725 with the U.S. Department of Energy. The publisher acknowledges the US government license to provide public access under the DOE Public Access Plan (http://energy.gov/downloads/doe-public-access-plan).

The growth of high-performance computing (HPC) now makes it practical to conduct detailed ACOPF studies across thousands of generators, evaluating each generator's removal against reliability, system cost, and fuel mix metrics. The Exascale Grid Optimization (ExaGO) toolkit [13] – developed at Oak Ridge National Laboratory and first deployed on the Frontier exascale supercomputer in 2023 [14] – provides an open-source platform capable of solving large-scale stochastic, security-constrained, and multi-period ACOPF problems on parallel and heterogeneous architectures, and has been validated on networks comparable in scale to real-world interconnections.

This paper demonstrates the approach on the 70,000-bus synthetic Eastern Interconnection model, developed at Texas A&M University (TAMU) [15], ranking over 8,000 operating generators by their system impact and identifying those whose retirement or deferral is most consequential. We also discuss solution time, the sensitivity of results to solver starting conditions, and the limitations of ACOPF in non-convex solution spaces.

This study is organized as follows. Section II describes the simulation design and scenario structure. Section III presents results, including computational performance, convergence behavior, and sensitivity analysis. We summarize results and discus the next steps in section IV.

## II. Model Inputs and Simulation Design

### A. Model inputs

Our goal is to quickly identify which units are essential to the grid and which can potentially be retired or temporarily curtailed for resource-mix reasons. We are also interested in the consequences of such removal on system cost and fuel mix. We apply N-1 logic to each operating generator on the grid, evaluating grid performance and fuel mix in the absence of the generator in question. While we apply this to only a single peak hour, the same high-fidelity, physics-based modeling could be performed for every hour of the year to produce annual assessments.

The simulation logic is applied to the 70,000-bus TAMU case, which is comparable in scale to large real-world interconnections such as the U.S. Eastern Interconnection or the EU grid. The case has 10,390 generators, including 8,107 generators for which the status is 'on'. The online generators represent 78% of all units and 87% of the technical minimum ($P_{min}$) of the system. TAMU case units approximately follow the power distribution. Larger generating units can reach 1000-1400 MW in size, while smaller can be as small as 0.1 MW. The generation mix is dominated by natural gas (41.1%) and coal (33.1%), followed by nuclear (11.0%), hydro (5.4%), fuel oil (5.3%), and inverter-based resources (4.1%). Average unit size varies by fuel type: nuclear units average 1,025 MW, coal 288 MW, natural gas 96 MW, and other fuel types between 3-54 MW (Fig. 1). Further, in the TAMU data, natural gas, coal and nuclear units use quadratic fuel cost curves, fuel oil units use a constant cost function, and hydro and all IBR units have zero cost.

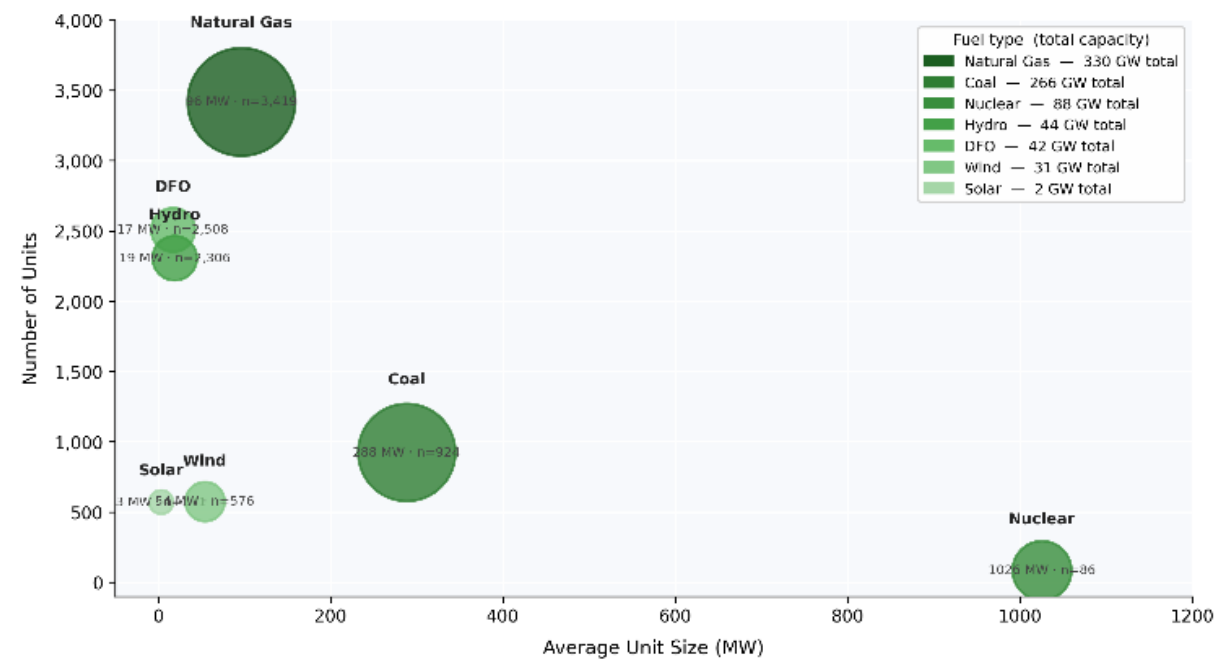

**Fig. 1.** TAMU 70,000 bus system generation mix.

The true extent of generator importance is related to grid constraints: removing even a small generator in a load pocket may be more difficult than removing a comparatively large generator in a capacity surplus part of the grid. The original grid has significant transmission overcapacity, so it is modified to reduce MVA limits to 150% of the power flow obtained in a baseline converged scenario (Fig.2).

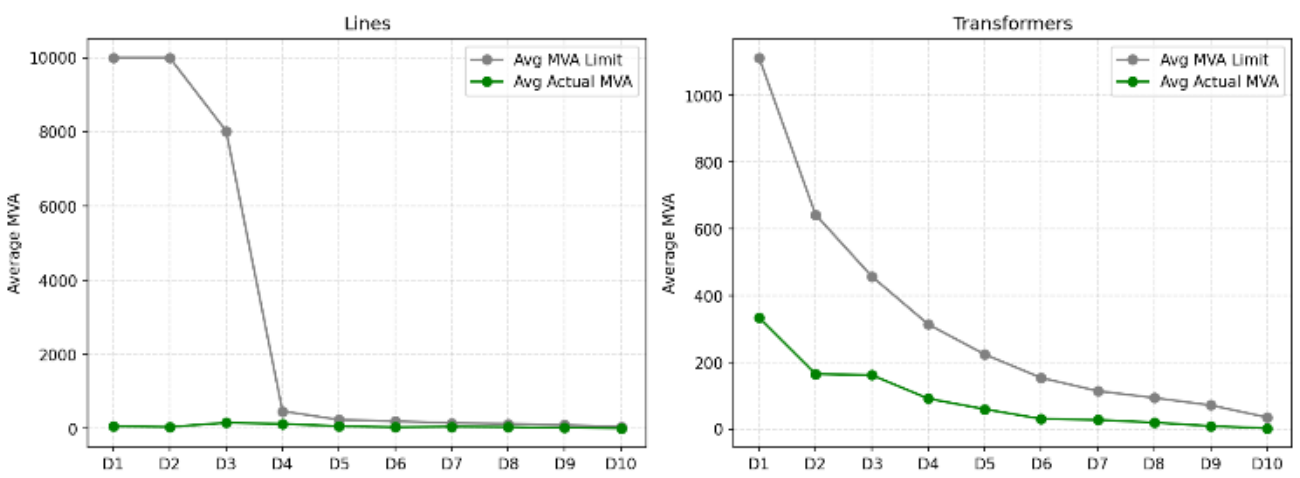

**Fig. 2.** TAMU 70,000 bus system branch loading (unlimited lines are defined using 9999 MVA limits, causing the first 30% of lines to show unusually high MVA values).

### B. Simulation design

Simulation scenarios are designed to (a) not disconnect any of the operating generators, (b) disconnect each of the 8,107 operating generators independently. For each of the generators, we assess (a) whether the grid can operate without that generator, and, if so, (b) what the economic and (c) fuel mix consequences are. The first question can be answered by observing the ACOPF convergence. The second question can be answered by evaluating the total system cost. Since each unit has a cost curve associated with it, we can compute the cost associated with loading each generator to a certain MW output. The third question can be answered by knowing the exact load of each unit and, subsequently, the total generation mix contributed by all units in a given fuel category.

Table I. Summary of model runs.

| Group id | Model | No line limit | Shed load | Slack variables | Analysis stage |
|---|---|---|---|---|---|
| From file, 100 iterations | ACOPF | 0 | 0 | 0 | Calibration (data of interest: percentage converged, system cost, total time) |
| From file, 200 iterations | ACOPF | 0 | 0 | 0 | |
| From file, 300 iterations | ACOPF | 0 | 0 | 0 | |
| From file, 400 iterations | ACOPF | 0 | 0 | 0 | |
| From file, 500 iterations | ACOPF | 0 | 0 | 0 | |
| Main | ACOPF | 0 | 0 | 0 | Ranking (percentage converged, system cost, unit MW loading, total time) |
| Shed Load | ACOPF | 0 | 1 | 0 | |
| Unlimited Network | ACOPF | 1 | 0 | 0 | |
| Main | DCOPF | 0 | 0 | 0 | Robustness checks (system cost, total time) |
| Shed Load | DCOPF | 0 | 1 | 0 | |
| Unlimited Network | DCOPF | 1 | 0 | 0 | |
| Flat start debug | ACOPF | 0 | 0 | 1 | |
| From file debug | ACOPF | 0 | 0 | 1 | |
| Midpoint debug | ACOPF | 0 | 0 | 1 | |
| Flat start | ACOPF | 0 | 0 | 0 | |
| From file | See Main | | | | |
| Midpoint | ACOPF | 0 | 0 | 0 | |

Table II. Descriptive statistics for main stage ACOPF runs.

| id_file | Main | Shed Load | Unlimited |
|---|---|---|---|
| convergence_% | 86.58 | 99.89 | 100.00 |
| cases converged | 7020 | 8099 | 8108 |
| Pg min [GW] | 610.8 | 608.4 | 610.2 |
| Pg mean [GW] | 610.8 | 610.8 | 610.3 |
| Pg max [GW] | 611.5 | 611.3 | 610.6 |
| Qg min [Gvar] | 105.4 | 103.2 | 90.6 |
| Qg mean [Gvar] | 106.0 | 104.1 | 92.2 |
| Qg max [Gvar] | 123.7 | 124.7 | 94.1 |
| Objective min [$10^7$] | 16.47 | 16.47 | 16.44 |
| Objective mean [$10^7$] | 16.48 | 16.49 | 16.44 |
| Objective max [$10^7$] | 16.64 | 18.96 | 16.52 |
| Solve time min [s] | 68.22 | 79.29 | 31.50 |
| Solve time mean [s] | 79.90 | 86.19 | 33.72 |
| Solve time max [s] | 173.46 | 213.80 | 39.92 |

Since the load and the grid conditions can affect the importance of a generator and the resulting fuel mix, we additionally run tests with load shedding allowed ("Load Shed" scenarios) and with unlimited MVA for transmission ("Unlimited Network" scenarios). We further apply several robustness checks. We first compare ACOPF solution with direct current optimal power flow (DCOPF) solution for each of the three groups. Given that ACOPF problem is not convex and may result in different solutions depending on starting conditions, we also provide different start types: flat start, midpoint, and "from file" (usually a solution from the most recent converged case). We run robustness checks both in regular run mode and debug mode. We also apply calibration based on a different number of solution steps. The summary of all main scenarios and robustness checks is provided in Table I. There was total of 121,620 scenarios, processed through 15 groups: 5 calibration runs followed by 3 production groups and 8 robustness checks. The descriptive statistics for the main stage outputs is provided in Table II.

## III. Results and Discussion

### A. Generator importance: reliability, cost, and fuel mix

We first review the results of the main stage runs. This includes the "Main" group of simulations that disconnects a generator without any further grid relief actions, plus the additional "Load Shed" and "Unconstrained" groups, for ACOPF. Results are summarized in Fig. 3.

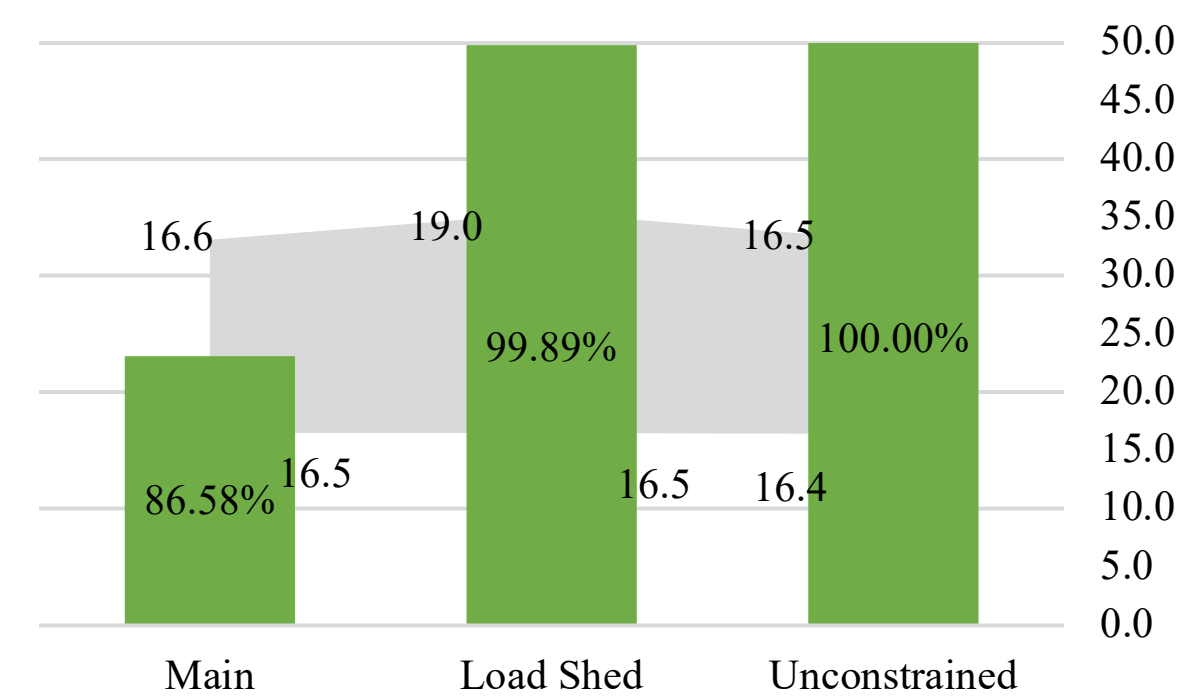

**Fig. 3.** Convergence results and system cost for main stage runs (left axis: percent converged cases, right axis: value of the objective function, $10^6$).

The convergence results show that about 87% of operating generators in the TAMU dataset can be retired or temporarily disconnected in N-1 logic, if necessary, subject to further reliability analysis (e.g. N-k contingencies, transient and stability etc.). The objective function value was evaluated only for converged cases and depends on the generator that is being disconnected. For the “Main” group, it ranges between 16,470,773.2 and 16,638,239.6. Interestingly, the baseline with no generator disconnected yields a cost of 16,472,308.2, which indicates that removing some generators (in this case, there were 6 such generators) can reduce system cost through redispatch to other units. The change in fuel mix is not particularly large, given that only one of 8,107 generators is being disconnected at a time. Shares in fuel mix across 8108 scenarios are reported in Table III.

Table III. Min/max share in fuel mix percentage.

| Fuel type | Main | Shed Load | Unlimited |
|---|---|---|---|
| natural gas | 50.10-51.20 | 49.96-51.20 | 51.19-51.96 |
| coal | 18.10-19.93 | 18.07-19.42 | 16.49-17.85 |
| nuclear | 14.11-14.84 | 14.14-14.57 | 14.52-15.40 |
| heating oil | 6.35-6.70 | 6.54-6.72 | 6.59-6.89 |
| hydro | 5.23-5.66 | 5.40-5.68 | 5.31-5.82 |
| inverter based | 3.69-4.01 | 3.79-4.02 | 3.67-4.12 |

Shedding load was helpful in most cases, as the number of solved cases increased from about 87% to over 99%. System cost, however, changed very significantly from scenario to scenario. The minimum cost was found to be 16,467,357.5 (only slightly below the baseline), but the maximum cost was as high as 18,959,426.9, due to the high cost of load shedding. The fuel mix changed mildly compared to the original.

The most impactful alternative scenario involved relaxing branch MVA limits. While unconstrained transmission is not physically realizable, this configuration illustrates the potential benefit of major transmission investment. The convergence reached 100% as any generator could be replaced by the system reserve. Furthermore, the system cost declined throughout most scenarios: the minimum was at 16,437,187.5, the maximum at 16,521,859.2. Only 93 scenarios were more expensive than the original baseline, usually, if a cheap base unit was removed. The fuel mix changed very significantly, indicating the potential value of grid for using different fuel.

With these results we create a priority list, ranking generators in terms of importance along any of the dimensions discussed above. Table IV shows top 10 generators that, if removed, either cause the system cost to decline, or to increase minimally. Most of these units appear consistently across all three run groups, indicating that their importance is robust to assumptions about load shedding and transmission constraints. The top 6 generators, if removed, improve the system cost across all three groups. The list is dominated by large coal units, though one large natural gas unit and several natural gas peaking units also appear. In real systems, such findings could carry significant implications for fuel mix policy and associated health impacts.

Table IV. Generators ranked by reliability (ascending), and system cost (descending). Retiring units 1-6 leads to system cost decrease; retiring units 7-10 causes small cost increase. No blackout risks were identified for any of the 10 cases.

| rank | Main | Load Shed | Unconstrained |
|---|---|---|---|
| 1 | bus 69998 gen 1 (568 MW coal) | bus 69998 gen 1 (568 MW coal) | bus 61573 gen 1 (793.2 MW coal) |
| 2 | bus 69997 gen 1 (568 MW coal) | bus 61573 gen 1 (793.2 MW coal) | bus 69998 gen 1 (568 MW coal) |
| 3 | bus 61573 gen 1 (793.2 MW coal) | bus 59813 gen 1 (558 MW coal) | bus 59813 gen 1 (558 MW coal) |
| 4 | bus 59813 gen 1 (558 MW coal) | bus 69997 gen 1 (568 MW coal) | bus 59934 gen 1 (900 MW coal) |
| 5 | bus 59934 gen 1 (900 MW coal) | bus 59934 gen 1 (900 MW coal) | bus 61081 gen 1 (653.3 MW gas) |
| 6 | bus 61081 gen 1 (653.3 MW gas) | bus 61081 gen 1 (653.3 MW gas) | bus 69997 gen 1 (568 MW coal) |
| 7 | bus 44081 gen 2 (0.45 MW gas) | bus 44081 gen 2 (0.45 MW gas) | bus 65849 gen 1 (789 MW coal) |
| 8 | bus 44405 gen 2 (0.5 MW gas) | bus 44405 gen 2 (0.5 MW gas) | bus 61856 gen 1 (567.9 MW coal) |
| 9 | bus 44504 gen 1 (0.6 MW gas) | bus 44504 gen 1 (0.6 MW gas) | bus 44081 gen 2 (0.45 MW gas) |
| 10 | bus 67556 gen 1 (0.9 MW gas) | bus 67556 gen 1 (0.9 MW gas) | bus 44405 gen 2 (0.5 MW gas) |

Solution time is of separate interest, as HPC can help quickly set up and calibrate different generation portfolio scenarios. Total computation time per group ranged from 1.5 to 6 minutes across 8,108 scenarios, using 427 CPU nodes. This hardware setup is larger than a regular corporate cluster, but is reasonable for large utility computational resources and for cloud servers. Individual scenarios in the “Main” group took on average 92.98 seconds, including scenarios that did not converge after maximum number of solver iterations was reached. The computations that involved shedding load resulted in an average of 86.34 seconds per scenario and 4 minutes 40 seconds per group. The total computational time was lowest for “Unlimited Network” scenarios, with about 33.71 seconds per scenario and a total group time of about 1.5 minutes. For a large utility or transmission system operator, computation times would differ in two offsetting ways. Fewer available nodes would slow execution, but smaller system sizes would accelerate it. The TAMU case had 70000 buses, while PJM, for instance, has about 25000. Computational time depends on the starting conditions, which is further discussed in the robustness checks.

### *B. Robustness checks*

To assess the sensitivity of the results to alternative models or simulation setups, we compare ACOPF with DCOPF analyses and examine the robustness of results to initial values and solver configuration. The difference between ACOPF and DCOPF shows how far the solution is from the "true optimum" in the presence of reactive power constraints. ACOPF objective function values were predictably higher than those of DCOPF. In over 99% of cases across all three groups, the ACOPF solution was found within 10% of the corresponding DCOPF solution, with differences tightly concentrated around the mean: 91.1-99.6% of observations lay within one standard deviation (Fig. 4).

DCOPF was also found to solve faster, with group-level runtimes of 34 seconds to 2 minutes 30 seconds, 2-3x faster than ACOPF. The modest runtime tradeoff indicates that it is possible to keep transmission constrained nodal ACOPF as the preferred formulation, as its accuracy outweighs the computational costs.

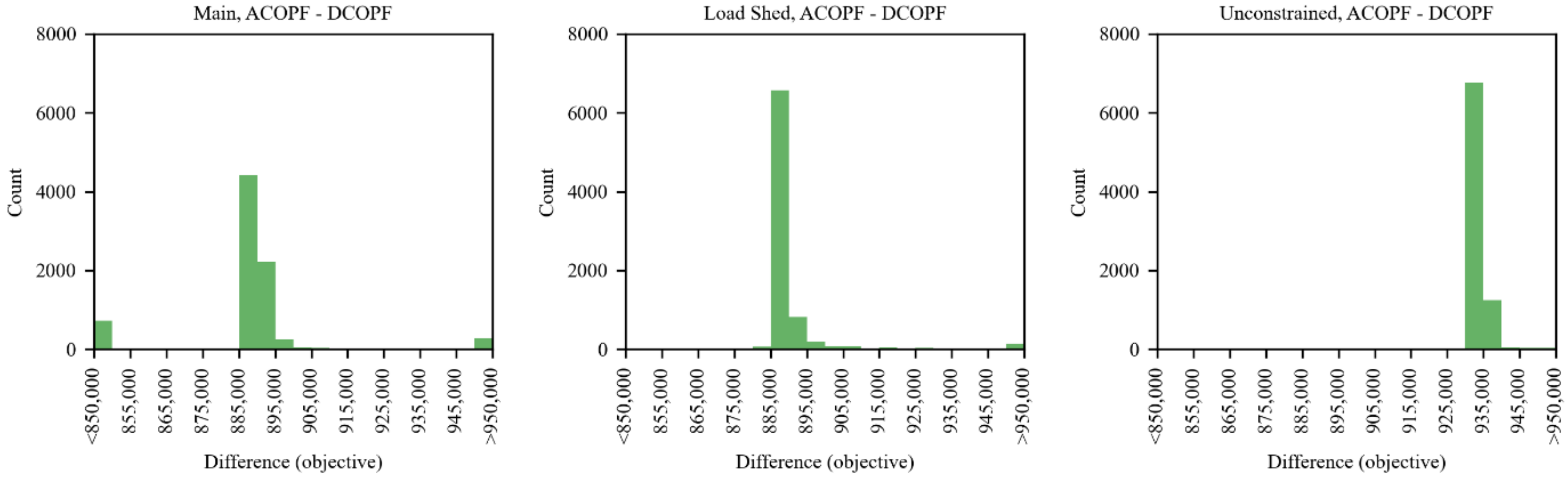

**Fig. 4.** Comparison of objective functions for ACOPF and DCOPF for each production run.

Table V. Robustness checks for generator ranking. Generators ranked for DCOPF runs on reliability, ascending, and system cost, descending. The original colors from Table III preserved to show the change in the order of generators.

| rank | Main | Load Shed | Unconstrained |
|---|---|---|---|
| 1 | bus 61573 generator 1 | bus 61573 generator 1 | bus 35471 generator 1 |
| 2 | bus 27051 generator 1 | bus 27051 generator 1 | bus 35410 generator 1 |
| 3 | bus 21528 generator 1 | bus 21528 generator 1 | bus 65849 generator 1 |
| 4 | bus 69998 generator 1 | bus 69998 generator 1 | bus 16022 generator 1 |
| 5 | bus 21530 generator 1 | bus 21530 generator 1 | bus 59813 generator 1 |
| 6 | bus 65849 generator 1 | bus 59813 generator 1 | bus 59934 generator 1 |
| 7 | bus 69997 generator 1 | bus 65849 generator 1 | bus 61573 generator 1 |
| 8 | bus 59707 generator 1 | bus 69997 generator 1 | bus 37081 generator 1 |
| 9 | bus 61856 generator 1 | bus 21529 generator 1 | bus 69998 generator 1 |
| 10 | bus 59934 generator 1 | bus 59934 generator 1 | bus 65466 generator 1 |

The next question is the effect of initial values. Because ACOPF is a non-convex problem, initial values can influence both the convergence path and the final solution. We compare three initialization strategies: flat start, midpoint, and "from file", on the system cost and solution time. The flat start and midpoint trials took significantly longer time to converge. To ensure comparability while preserving computational resources, all three groups were rerun in debug mode with slack variables enabled. Slack variables are active and reactive power injections or sinks that would not exist in the original power flow case, but allow to balance power flow at a model cost of $10,000 per MW or Mvar. The use of slack variables ensured convergence, although at a higher computational time and objective cost (Fig. 5).

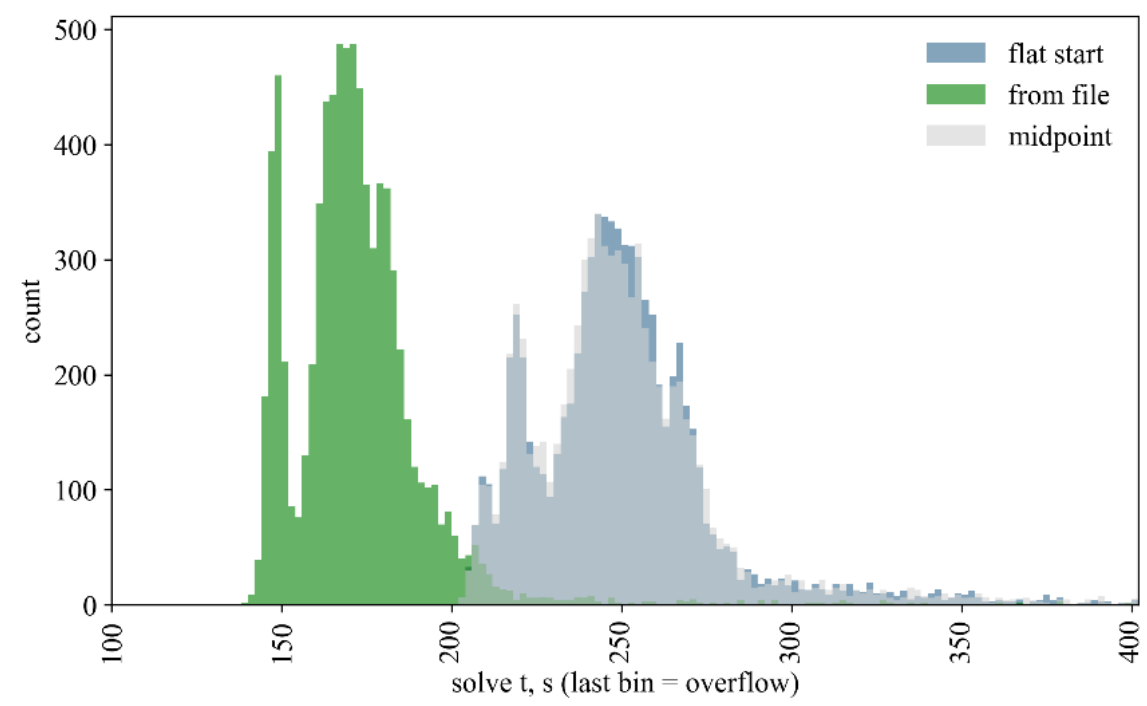

**Fig. 5.** Comparison of runtime for each starting condition.

Objective function values were effectively identical across starting conditions, differing by at most 0.048% on average and diverging in only 18 scenarios. This indicates no accuracy advantage from any particular initialization strategy. The Results were, however, different from the computational runtime viewpoint. Flat start and midpoint runs required almost double the time, which confirms that "from file" is a more efficient approach for the main runs.

It is also important to discuss optimization solver settings. The number of solver iterations has a significant effect on runtime but, beyond a threshold, yields no improvement in solution quality. We calibrated this threshold by running trial cases at step counts from 100 to 500 in increments of 100. The number of converged cases started with 83.0% at 100 steps and plateaued at 86.57%, while the objective function value for converged cases remained unchanged to within 0.01% across all step counts. We therefore adopt 500 steps as the optimal setting, beyond which additional iterations offer no improvement in convergence rate or solution quality.

## IV. Conclusions and Future Work

This paper presents an HPC-based framework for rapidly assessing the importance, operating cost impact, and fuel mix consequences of individual generating units across an entire system. We evaluated over 8,000 generators in the 70,000-bus TAMU synthetic grid under multiple operating scenarios in just under six minutes per run. This demonstrates that comprehensive generation portfolio assessments, historically time-consuming and infrequently performed, can be conducted routinely and at low computational cost. While the present study focuses on a single peak hour, the same approach can be extended to every hour of the year (provided sufficient compute resources), enabling full annual assessments of generator importance and retirement impacts.